# Controlling Carbon Nanostructure Synthesis in Thermal Plasma Jet: Correlation of Process Parameters, Plasma Characteristics, and Product Morphology

Taki Aissou[a], Jérôme Menneveux[a], Fanny Casteignau[a,b], Nadi Braidy[a,b], Jocelyn Veilleux[a]

[a] *Plasmas, Processes & Integration of Nanomaterials (2PIN) Laboratory, Department of Chemical Engineering and Biotechnological Engineering, Université de Sherbrooke, 2500 Boulevard de l'Université, Sherbrooke J1K 2R1, QC, Canada*

[b] *Institut Interdisciplinaire d'Innovation Technologique (3IT), Université de Sherbrooke, 3000 Boulevard de l'Université, Sherbrooke J1K 0A5, QC, Canada*

## Abstract

Thermal plasma has emerged as an effective approach for producing carbon nanostructures without the need for specific catalysts nor substrates. While efforts have focused on the effect of process parameters such as reaction pressure, input power or carbon source, the intricate role and relationship with plasma characteristics like density and temperature are often overlooked due to the complexity of the environment. This study addresses this gap by establishing a correlation between process parameters, plasma characteristics, and product morphology, essential for controlling the growth of carbon nanostructures. We explored the impact of carbon precursor type ($CH_4$ and $C_2H_2$), hydrogen, pressure, and flow rate on nanostructure formation. Using *in situ* optical emission spectroscopy (OES), we mapped the distribution of both temperature and dicarbon molecule ($C_2$) density within the plasma jet. We demonstrate that the growth of low-density nanostructures, such as carbon nanohorns (CNHs), is favoured at dilute $C_2$ local densities and high temperatures, while denser nanostructures, such as onion-like polyhedral graphitic nanocapsules (GNCs), are favoured at higher $C_2$ densities and lower temperatures. The carbon density can be controlled by the flow rate and the pressure, which in turn significantly influence the nanostructure morphology, evolving from graphene nanoflakes (GNFs) to GNCs as either parameter increases. Increasing the $H/C$ ratio from 1 to 8 resulted in a morphological transition from CNHs to GNFs. During the synthesis, the plasma jet temperature surpassed 3,000 K, with crystalline growth occurring 50 to 100 mm below the nozzle.

# 1 Introduction

Research on nanomaterials has been the subject of increasing interest and is now one of the fastest-growing areas. Carbon allotropes, including graphene [1], nanotubes [2], nanohorns [3], and fullerenes [4] are endowed with inimitable optical, electrical, mechanical, and chemical properties [5]. The remarkable emergence of these members of the carbon family has opened up new possibilities and perspectives for their integration into various industrial applications such as nanoelectronics [6], energy storage [7], drug delivery systems [8], and biosensors [9].

Carbon bottom-up synthesis methods are divided into low/medium and high-temperature approaches. Low/medium temperature methods like CVD [10], PECVD [11], epitaxy [12], combustion [13], and pyrolysis [14] produce high-purity materials but suffer from slow reaction kinetics and low yield, restricting commercial and large-scale use. High-temperature methods involve sublimation (solid, e.g., graphite), vaporization (liquid, e.g. alcohols), or dissociation and ionization (gas and plasma, e.g. hydrocarbons). These methods, including arc discharge [15] and laser ablation [16], produce highly crystalline nanoparticles but are often limited by their electrode lifetimes, production rates or contamination [17].

Thermal plasma jet processes like direct current (DC) plasma torch [18], microwave (MW) plasma torch [19], and radiofrequency (RF) plasma torch [20] have become increasingly prevalent as alternatives. These methods use hydrocarbon gas precursors that require less energy for decomposition and offer advantages like increased productivity, energy efficiency, high conversion rates, wide range of operating pressures, and process simplicity [21]. In particular, the RF plasma torch stands out due to its electrodeless operation and good yield, owing to its large volume and stream velocity. Historically, RF plasma torches have been used to achieve the synthesis and the commercialization of microdiamonds (1987) [22], fullerenes (2001) [23], nanotubes (2007) [2], graphene nanoflakes (2010) [24], nanohorns (2022) [3] and onion-like graphitic nanocapsules (2023) [25] (see Figure 1).

In the context of high-temperature environments, optimization of process parameters relating to plasma growth of carbon nanostructures is often based on post-synthesis observation using characterization tools such as transmission electron microscopy (TEM), scanning electron microscopy (SEM), Raman spectroscopy, etc. The extreme processing conditions of the plasma jet, including high temperatures and non-atmospheric pressures, make direct observation of the nucleation and growth processes challenging. Given this limitation, efforts devoted to synthesis attempts have mainly focused on correlating process parameters (e.g., precursor type [26], pressure [25], power [27]) with the various carbon nanostructures that can be obtained. In contrast, less attention has been given to studying the plasma characteristics (e.g., temperature, species density) and the factors responsible for the formation of specific carbon morphologies. Hence, the production of carbon nanostructures is often achieved empirically, without an exhaustive understanding of the underlying nucleation and growth mechanisms. To understand these mechanisms, it is essential to investigate the plasma behaviour during synthesis, which requires the identification of species (atoms, ions and molecules), species densities, species spatial distributions, plasma temperature profiles, quenching rates, residence times, and so on. Collectively, these parameters create the nucleation and growth environment of the possible carbon nanostructures in the plasma jet.

Some theoretical studies have investigated the thermal plasma synthesis of carbon nanostructures, such as graphene nanoflakes [28] and nanotubes [29], and have led to a better understanding of the mechanisms behind carbon nanostructure formation. These studies have integrated the thermodynamics and kinetics of carbon nucleation and growth, while modelling the flow dynamics in the plasma and predicting the temperature and velocity distributions. It has been reported that nucleation typically occurs between 3,000 and 5,000 K, and is followed by growth, which requires a sufficient residence time within a plasma jet [28,29]. However, existing numerical models developed remain approximate and must be validated by *in situ* measurements during nucleation and growth. These processes manifest at atomic and molecular scales and are often oversimplified by continuous approaches such as computational fluid dynamics (CFD) models and aerosol dynamics [28,29].

An effective approach to probing plasma involves the use of optical emission spectroscopy (OES), a recognized non-invasive technique for characterizing plasma. While OES has been previously applied in numerous studies during the synthesis of carbon nanomaterials through methods like CVD [30], laser [31,32], or arc discharge [33], its application within thermal plasma jet processes has been relatively scarce, pursued only by a handful of research groups [34–36]. OES performed during thermal plasma synthesis of GNFs have confirmed that the presence of dicarbon molecule ($C_2$) in the plasma jet is correlated with the production of GNFs [36], as expected by thermodynamic calculations [37]. However, these observations were limited to a specific localized point of the plasma jet which exceeds 200 mm in length, therefore neglecting the spatial distribution of species within the plasma. Furthermore, the study did not draw conclusive findings regarding the nucleation zone nor the associated temperature within the plasma jet. Considering the central role played by $C_2$ in the formation of carbon nanostructures, mapping the $C_2$ density and temperature in the plasma jet by OES should bring further insight into the complex mechanisms at play [38].

The purpose of this study is to synthesize and conduct a comprehensive investigation of carbon nanostructures produced by the decomposition of $CH_4$ and $C_2H_2$ in thermal plasma jets. In this work, the influence of the different growth conditions of carbon nanostructures is examined by adjusting the plasma operating parameters. The synthesis process is analyzed with *in situ* spatially resolved OES. The morphological and structural qualities of the synthesized carbon nanostructures are analyzed by transmission electron microscopy (TEM) and Raman spectroscopy. The relationship between process parameters, plasma characteristics and product morphology is then discussed.

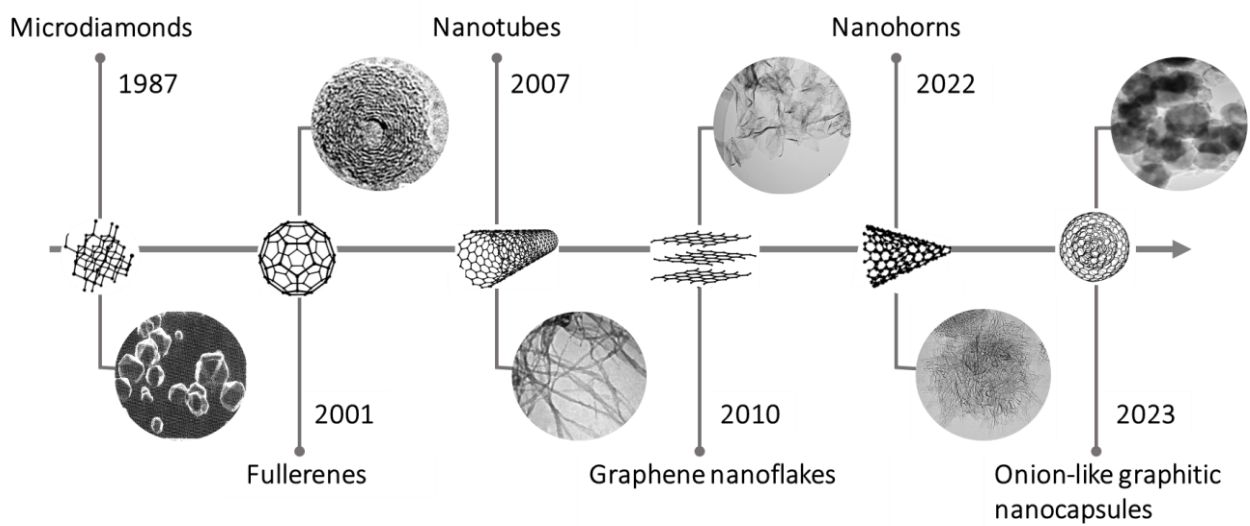

Figure 1. The synthesis history of the carbon nanostructures using inductively coupled plasma torch: microdiamonds [17], fullerenes [23], nanotubes [2], graphene nanoflakes [24], nanohorns [3], onion-like nanocapsules [25].

## 2 Experimental methods

### 2.1 Synthesis of carbon nanomaterials

Figure 2 (a) details the schematic of the experimental setup used for carbon nanopowder synthesis. It consists of a radiofrequency inductively-coupled plasma (ICP) torch (PL-50, TEKNA Plasma System Inc. Canada), mounted on top of a main synthesis chamber maintained under vacuum and connected to an auxiliary chamber equipped with a filtration system. The ICP torch, powered by a generator delivering up to 60 kW at 3 MHz, has a water-cooled ceramic tube encircled by a copper coil. For torch operation, two argon gas streams are introduced in the plasma torch: central gas and sheath gas [39]. In this work, a precursor gas probe, aligned with the coil area, introduces the carbon precursor and, in some cases hydrogen, to investigate the effect of the $H/C$ ratio.

To determine the zone of the plasma jet where crystallization occurs, we collected the carbon deposited at different distances from the torch using a stainless-steel substrate. Figure 2 (b) shows the horizontal cylindrical reactor that was used to collect deposits on said substrate, employing parameters similar to those for the synthesis reactor in Figure 2 (a).

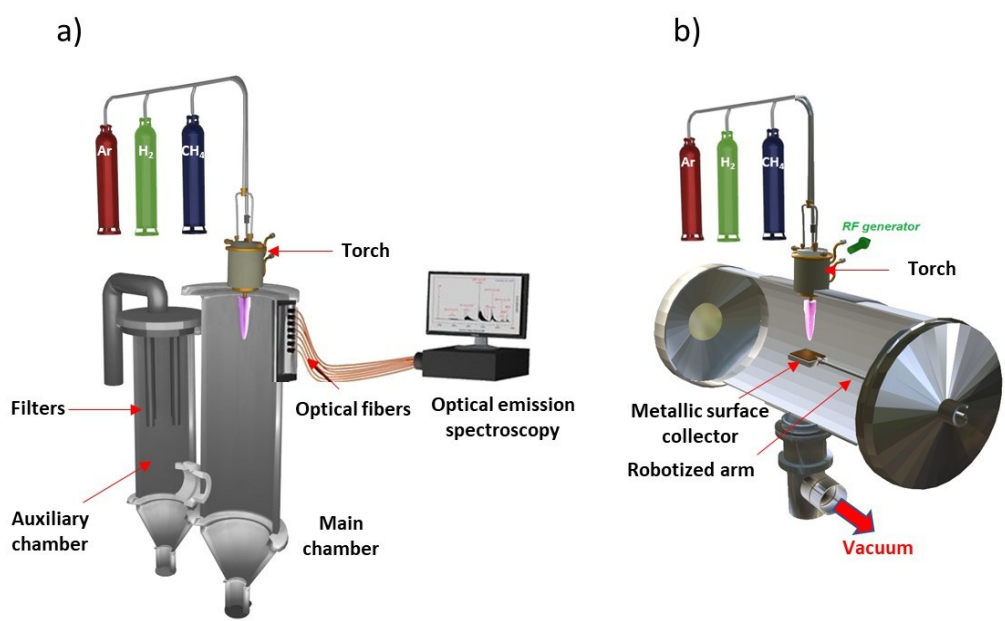

Figure 2. Schematic illustration of the experimental equipment.

A series of experiments were carried out in argon plasma using $CH_4$, $C_2H_2$, and $H_2$, operated under various pressures and different precursor gas flow rates (

Table 1). No powder was produced for pressures < 13 kPa, $H/C$ ratios > 10, or hydrocarbon flow rates < 0.5 slpm.

Table 1. Plasma parameters for carbon nanopowder synthesis.

| Process parameters | Value |
|---|---|
| Sheath gas (Ar, slpm) | 60 |
| Central gas (Ar, slpm) | 15 |
| Precursor ($CH_4$ or $C_2H_2$, slpm) | 0.5, 0.75, 1.5, 3, 4 |
| $H_2$ (slpm) | 1.5-10.75 |
| Rector pressure (kPa) | 13, 25, 40, 45, 70 |
| Power (kW) | 20 |

## 2.2 In situ plasma diagnostics

OES measurements were made on the plasma jet to identify the chemical species and map $C_2$ relative density and temperature. The OES measurement was performed under different synthesis conditions with an IsoPlane SCT-320 spectrometer (Princeton Instruments) connected to a PIXIS:256E CCD camera. The 1800 g/mm grating allowed a resolution of 0.037 nm and covered a spectral range of 38 nm. A 400 nm to 770 nm window was selected to observe the emission of Ar, C, H, $C_2$, and CH atoms and molecules. To record the complete sequence, a "Step and Glue" function was used. To carry out the OES measurements, light from the plasma was collected using seven optical fibres vertically spaced 15 mm apart. The upper fibre is located at the same level as the torch nozzle exit. The fibres were moved horizontally during the experiment to map the plasma. The OES maps were made in two steps: first, a 20 mm horizontal scan with a 2 mm step covering a vertical distance of -90 mm from the torch exit, then another scan covering the vertical distance from -105 mm to -195 mm. Figure 3 (a) shows a photo of a pure Ar plasma jet. Figure 3 (b) shows the typical optical emission spectrum of Ar plasma at 45 kPa, obtained using one of the optical fibres installed to capture the light and thus collect the emission spectra. Figure 3 (c) shows a 2D OES map highlighting the intensity of the Ar I atomic line. In the Supplementary Information section (Suppl Info), we provide a comprehensive analysis and detailed calculations required to achieve both the temperature and species density maps observed within the plasma.

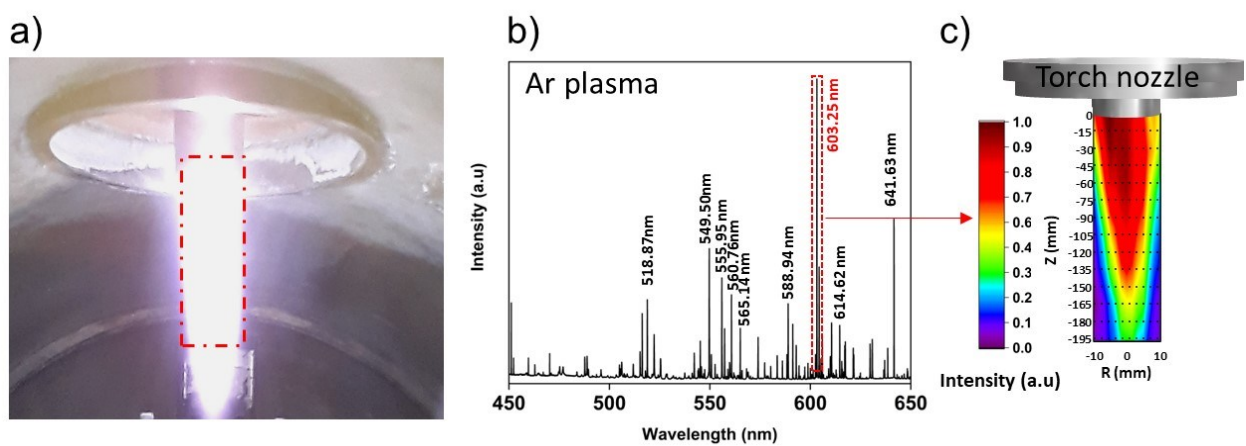

Figure 3. (a) Illustrative photograph of Ar plasma jet. (b) Typical optical emission spectrum of Ar plasma at 45 kPa. (c) 2D OES mapping showing the intensity of the Ar I (603.25 nm) atomic line. The vertical axis, Z, indicates the position of optical fibres relative to the plasma torch nozzle exit, while R denotes the radial distance from the center of the nozzle. Points on the map mark the positions where the optical fibres captured data from the plasma.

## 2.3 Ex situ nanomaterials characterization

After each synthesis using the reactor in Figure 2 (a), the black carbon nanopowder deposited on the inner walls and filters was collected and analyzed. The morphology and microstructure of carbon nanopowder were characterized by transmission electron microscopy (TEM, Hitachi H-7500) with an accelerating voltage of 120 kV and, for some samples, by high-resolution HRTEM (JEOL ARM 200F) equipped with a cold FEG electron source operated at 80 kV and an aberration corrector on the objective lens [40]. The structural quality of the carbon nanopowder was probed by Raman spectroscopy (Princeton Instruments, Acton SP-2500i) using a 30 mW, 514 nm wavelength laser.

## 2.4 Chemical reaction analysis

During the synthesis, the reactor chamber was connected to an online mass spectrometer (MS, *Edwards Vacuum*) for real-time analysis of gaseous species. The gas composition for stable reaction products such as $CH_4$, $C_2H_2$, $H_2$, and $C_4H_2$ was analyzed by the quadrupole mass detector. The mass spectrometer was connected to the vacuum pumping line, downstream the filters. Additionally, FactSage 8.2 software was employed to perform thermodynamic calculations for identifying potential species and molecules, predicting the optimal nucleation temperature, and discerning the influence of process parameters. This code, based on the minimization of the Gibbs free energy, enabled the analysis of various phases and components in equilibrium. The FactPS database covers all typical plasma species relevant to this study.

# 3 Results

Figure 4 shows the various morphologies of carbon nanostructures synthesized in an ICP thermal plasma environment. Depending on the conditions used, three types of carbon morphologies are generated: graphene nanoflakes (GNFs), carbon nanohorns (CNHs) and graphitic nanocapsules (GNCs). Table 2 summarizes the results of several 20-minute synthesis experiments, varying parameters such as precursor gas ($CH_4$, $C_2H_2$), flow rate, pressure, and the $H/C$ ratio. Table 2 also details the resulting structures, as identified by TEM and Raman spectroscopy, along with the powder production rate.

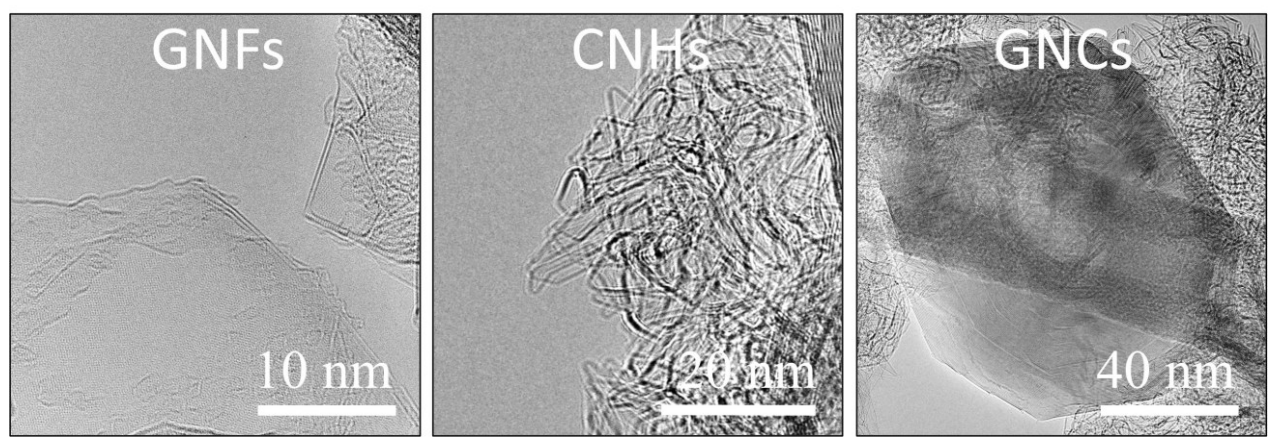

Figure 4. HRTEM images of graphene nanoflakes (GNFs), carbon nanohorns (CNHs) and graphitic nanocapsules (GNCs) synthesized using ICP torch.

Table 2. Table summarizing Raman spectroscopy results, TEM-identified structures, and production rates for samples produced under different synthesis conditions.

| Precursor | Pressure (kPa) | Precursor flow rate (slpm) | Hydrogen (H$_2$) | H/C ratio | Raman | | TEM-identified structures | Production rate (g/h) |
|---|---|---|---|---|---|---|---|---|
| | | | | | $I_D/I_G$ | $I_{2D}/I_G$ | | |
| CH$_4$ | 13 | 1.5 | No | 4 | N/A | N/A | N/A | 0 |
| | 25 | | No | 4 | 0.50 | 0.55 | GNFs | N/A |
| | 40 | | No | 4 | 0.53 | 0.52 | GNFs | 15 |
| | 45 | | No | 4 | 0.53 | 0.52 | GNFs | 17 |
| | | | Yes | 6 | 0.31 | 0.62 | GNFs | 12 |
| | | | Yes | 8 | 0.36 | 0.56 | GNFs | N/A |
| | | 3 | Yes | 6 | 0.63 | 0.43 | GNFs +GNCs | 20 |
| | | 4 | Yes | 6 | 0.76 | 0.35 | GNFs + GNCs | N/A |
| | 70 | 1.5 | Yes | 6 | 0.86 | 0.27 | GNFs + GNCs | 22 |
| C$_2$H$_2$ | 25 | 0.7 | No | 1 | 0.9 | 0.25 | CNHs + GNFs | N/A |
| | 40 | | No | 1 | 1.17 | 0.35 | CNHs | 18 |
| | 45 | | No | 1 | 1.16 | 0.35 | CNHs | N/A |
| | | | Yes | 6 | 0.32 | 0.62 | GNFs | 10 |
| | | | Yes | 8 | 0.35 | 0.6 | GNFs | N/A |
| | | 1.5 | Yes | 6 | 0.64 | 0.42 | GNFs | 20 |
| | | 2 | Yes | 6 | 0.75 | 0.37 | GNFs + GNCs | 20 |
| | 70 | 0.7 | Yes | 6 | 1.22 | 0.33 | Carbon black | N/A |

## 3.1 Carbon nanostructures characterization

### 3.1.1 Effect of carbon source and reaction pressure

The precursor type and the synthesis pressure both played a crucial role in the selectivity of the carbon nanostructure morphology (Figure 5). GNFs are synthesized with CH$_4$ at a pressure of 25 kPa. These structures are 100-200 nm large and are made of graphene stacks with less than 10 layers thick, judging from the edge thickness measured by TEM. With C$_2$H$_2$ at the same pressure, spherical carbon aggregates 25-50 nm in diameter composed of short tubules (5 nm long) are produced. These are known as seed-like and bud-like CNHs [3]. A small fraction of GNFs are systematically observed alongside the CNHs. At 40 kPa, GNFs are still produced using CH$_4$, and CNHs structures are dominant with C$_2$H$_2$ with no trace of GNFs. By further increasing the pressure to 70 kPa and using CH$_4$, graphitic nanocapsules (GNCs) with polygonized facets having sizes between 70 and 300 nm are obtained, while cage-like carbon black structures referred to as Ketjenblack® [41,42] (trademark registered to AKZO NOBEL CHEMICALS B.V.) are produced with C$_2$H$_2$.

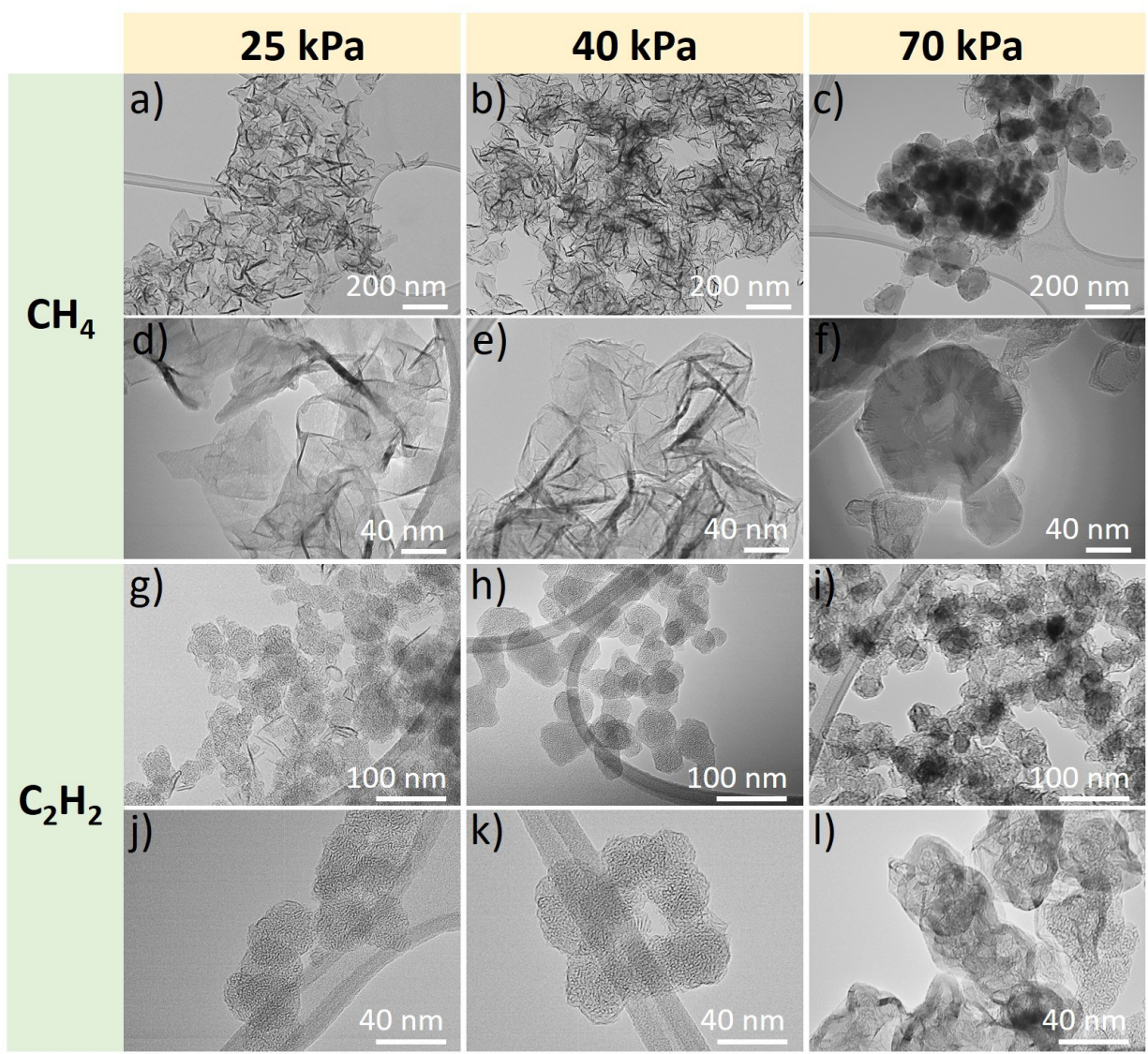

Figure 5. TEM images at different magnifications of carbon nanomaterials synthesized using 1.5 slpm of CH₄ at pressures of 25 kPa (a, d), 40 kPa (b, e), and 70 kPa (c, f), as well as those synthesized using 1.5 slpm of C₂H₂ at pressures of 25 kPa (g, j), 40 kPa (h, k), and 70 kPa (i, l).

### 3.1.2 Effect of $H/C$ ratio

The addition of H₂ dilutes carbon gas precursors and thus, the powder production rate. Figure 6 shows typical TEM images of nanostructures synthesized with CH₄ and C₂H₂ at several $H/C$ ratios. Here, the flow rate for C₂H₂ is set at 0.7 slpm, i.e. half the 1.5 slpm flow rate for CH₄, which guarantees an equivalent carbon contribution to the plasma jet for both precursors (the lowest possible $H/C$ ratio with CH₄ is 4. GNFs are exclusively produced with CH₄ at $H/C$ ratios of 4, 6 and 8. Conversely, using C₂H₂ at $H/C$ ratio of 1 results in the synthesis of seed-like and bud-like CNHs. The addition of H₂ to C₂H₂ favours GNFs rather than CNHs. GNFs produced from CH₄ and C₂H₂ at $H/C$ ratios of 6 and 8 exhibit similar morphological features.

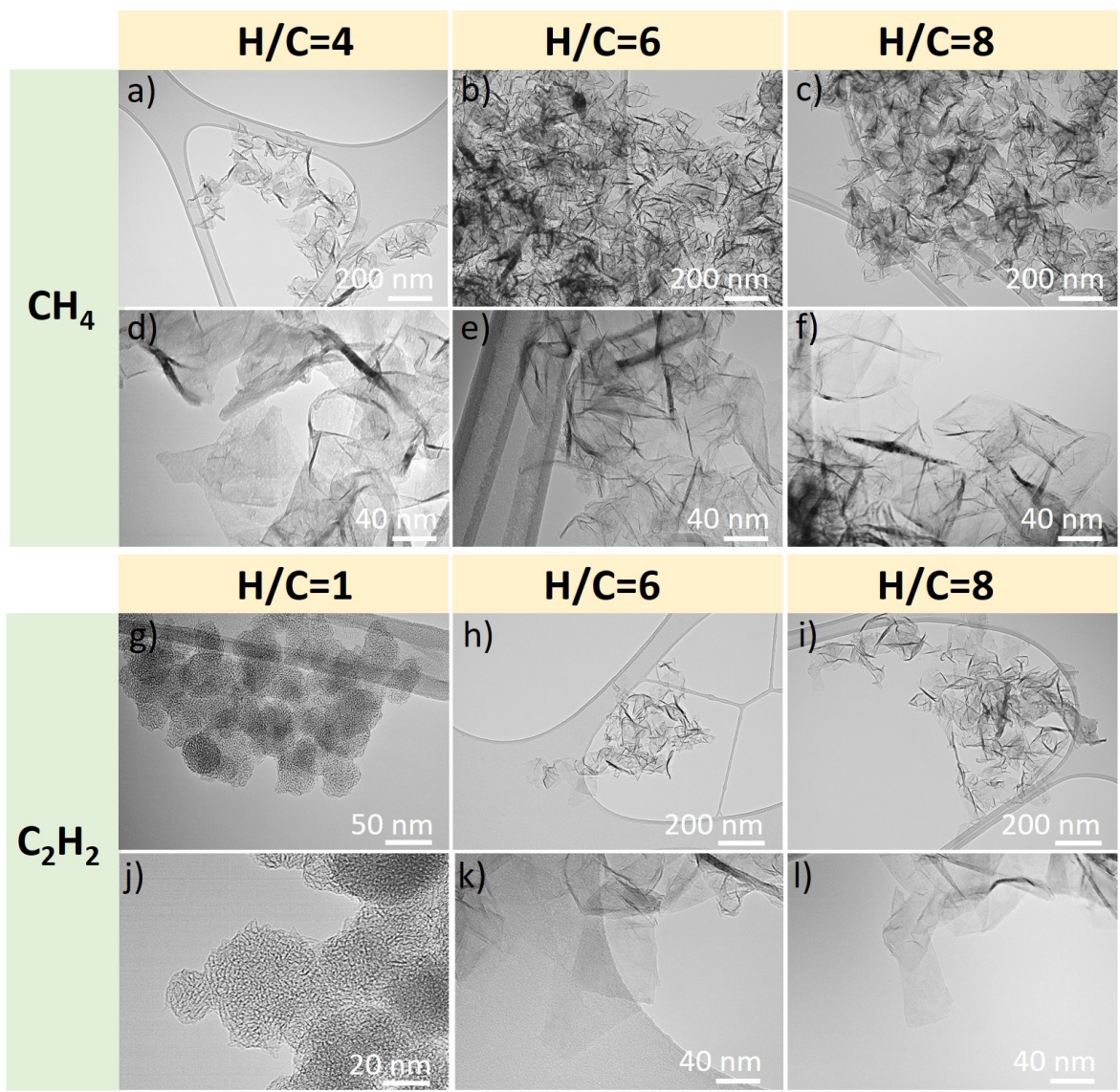

Figure 6. TEM images at different magnifications of the carbon nanomaterials prepared at 45 kPa using CH$_4$ (1.5 slpm) and C$_2$H$_2$ (0.7 slpm) at several $H/C$ ratios.

When employing $H/C > 1$, whether using CH$_4$ or C$_2$H$_2$, the resulting structures consistently manifest as GNFs with apparently identical morphologies. To deepen the analysis and discern subtle distinctions in the structure and quality of these GNFs, we turned to Raman spectroscopy (Figure 7). The three typical peaks of sp$^2$ nanocarbon appear in all Raman spectra and correspond to the D (1350 cm$^{-1}$), G (1580 cm$^{-1}$), and 2D (2700 cm$^{-1}$) bands. The D-band is attributed to a structural disorder or defective structure as well as to the presence of small graphite crystallites. G-band represents the E$_{2g}$ vibration of the graphite plane in sp$^2$ graphitic materials, and the 2D band corresponds to the second order of the D-band. Weaker features of the Raman spectra such as the D', D+D", and D+D' bands were also isolated by deconvolution but were not used in the analysis (Section 2.2, "Raman" in the Suppl Info.).

The intensity ratio $I_D/I_G$ is widely used to characterize the occurrence of defective carbon in graphitic materials: a lower value of $I_D/I_G$ generally implies a lower level of defects [43,44]. On the other hand, the intensity ratio of $I_{2D}/I_G$ is considered a criterion to compare the number of graphene layers: a higher value of $I_{2D}/I_G$ generally implies a lower number of graphene layers. The $I_D/I_G$ value of the carbon nanomaterials produced decreases with increasing $H/C$ ratio, indicating a reduction in defects and an improvement in crystallinity. Specifically, when using $C_2H_2$ as a precursor, the $I_D/I_G$ value decreased from 1.2 to 0.3 as the $H/C$ ratio increased from 1 to 6. Interestingly, the Raman spectrum of $C_2H_2$ ($H/C = 1$) associated with CNHs shows a high and broad D peak at a lower frequency, which can be attributed to (i) topological distortions resulting from layer folding, the presence of pentagon-heptagon defects including the pentagons required to generate the conical shape of the CNHs (Stone-Wales defects) [45,46]; and/or (ii) edges due to the finite length of the CNHs [47]. On the other hand, when using $CH_4$ as a precursor, the $I_D/I_G$ value decreased from 0.56 to 0.3 as the $H/C$ ratio increased from 4 to 6. The Raman spectrum of $CH_4$ ($H/C = 6$) shows a similar trend to that of $C_2H_2$ ($H/C = 6$), with a 2D band hinting that their products share similar structural characteristics. However, the $I_D/I_G$ ratio of samples produced with $H/C = 8$ is slightly larger than those produced with $H/C = 6$, suggesting the formation of additional structural defects in the GNFs. Thus, while hydrogen generally favours the formation of graphitic carbon with lower defect up to a certain $H/C$ ratio (in this case, up to 6), higher ratios can lead to increased defect formation.

TEM and Raman results indicate that a ratio $H/C = 6$ is favourable to the growth of GNFs with a relatively more ordered structure, fewer defects, and higher crystallinity. The results also indicate that the nature of the precursor is not the only factor determining the formation of GNFs and that the $H/C$ ratio in a plasma environment is the primary determining factor.

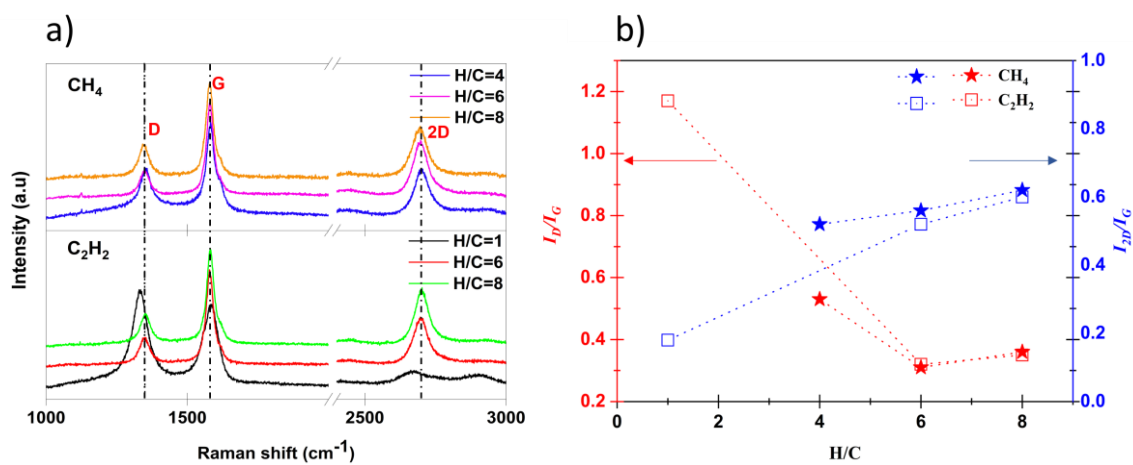

Figure 7. (a) Raman spectra of nanostructures prepared at 45 kPa using $CH_4$ (1.5 slpm) and $C_2H_2$ (0.7 slpm) at several $H/C$ ratio. (b) Raman peak ratios $I_D/I_G$ and $I_{2D}/I_G$ as a function of $H/C$ value.

### 3.1.3 Effect of hydrocarbon flow rate

In this section, the pressure and the power are kept constant at 45 kPa and 20 kW, respectively, and the $CH_4$ and $C_2H_2$ gas flow rates were increased to evaluate the impact of the carbon concentration in the plasma on the product morphology (Figure 8). For comparison purposes, the flow rate of $C_2H_2$ used here is half that of $CH_4$, which ensures an equal amount of C introduced in the plasma jet.

GNFs are exclusively formed with a flow rate of 1.5 slpm. Increasing to 3 slpm leads to thicker GNFs and the apparition of thin-shelled GNC. At a flow rate of 4 slpm, GNCs become dominant. The same trend was observed when increasing the flow rate using $C_2H_2$: a higher hydrocarbon flow rate in the plasma leads to denser carbon nanostructures, such as GNCs, rather than GNFs.

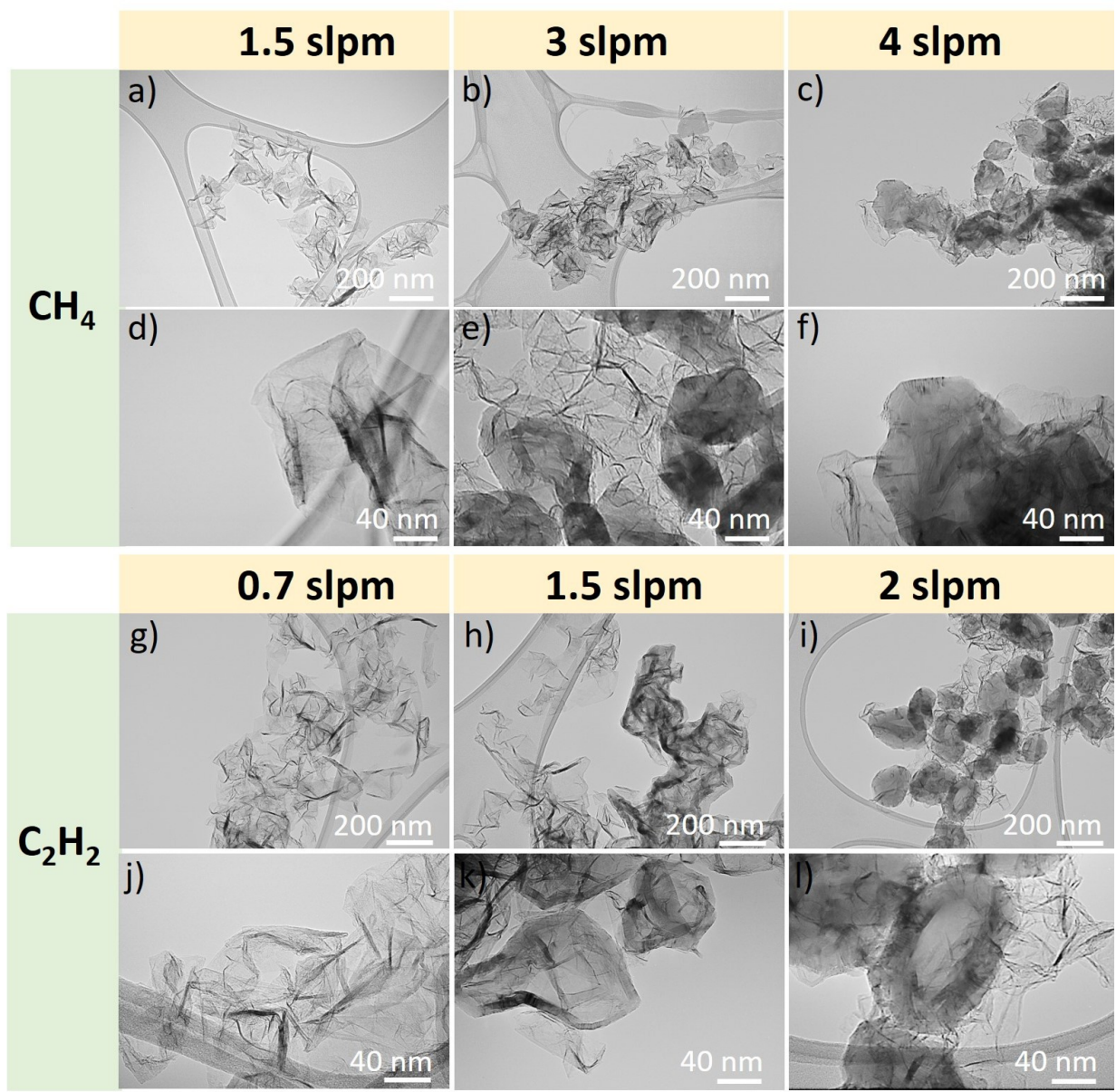

Figure 8. TEM images at different magnifications of the carbon nanomaterials synthesized at 45 kPa; using $CH_4$ + $H_2$ ($H/C = 6$) with flow rates of 1.5, 3, and 4 slpm; using $C_2H_2$ + $H_2$ ($H/C = 6$) with $C_2H_2$ flow rates of 0.7, 1.5, and 2 slpm.

## 3.2 Plasma diagnostics during the synthesis process

We demonstrated that the carbon source, reaction pressure, $H/C$ ratio, and hydrocarbon flow rate have a significant impact on the nanocarbon morphology produced in the plasma reactor. To understand the underlying mechanisms that dictate this morphology, it is imperative to analyze how these process parameters modulate the plasma intrinsic characteristics, namely temperature and relative species density. Numerous studies and empirical data identified the $C_2$ molecule as the building block for sp$^2$ carbon nanostructures [48–51]. To gain deeper insights
14

into the behaviour and distribution of C$_2$ within our system, we used OES to map both the temperature ($T_{C_2}$) and relative density of C$_2$ ($N_{C_2}$).

### 3.2.1 Effect of carbon source and reaction pressure

Figure 9 shows the 2D maps of $T_{C_2}$ and $N_{C_2}$ measured at 25 kPa, 40 kPa, and 70 kPa using CH$_4$ and C$_2$H$_2$. $T_{C_2}$ ranges from 3,000 K to 6,500 K over an axial distance of 195 mm. In all cases, $T_{C_2}$ was found to be higher in the off-axis region and slightly lower in the center. This phenomenon is attributed to the energy coupling in the ICP discharge, which is higher in the outer annular shell of the plasma. This toroidal or "donut" shape forms because of the induced electric field from the RF coil [52]. The electric field is most potent near the coil, decreasing towards the center, leading to intense heating in the outer regions. Consequently, the central region of the plasma experiences a reduced energy input resulting in lower temperatures in the central plasma region [52]. In addition, the central injection of carbon source gas also contributes to this observation. By comparing their $T_{C_2}$ values, C$_2$H$_2$/Ar plasma generally exhibits higher temperatures than CH$_4$/Ar plasma, particularly in the upper regions close to the torch. This observation aligns with a previous study where it was found that C$_2$H$_2$/Ar plasma was warmer than CH$_4$/Ar plasma [27]. On the other hand, the overall temperatures decrease lightly by increasing the pressure from 25 kPa to 70 kPa. The increase in pressure causes an increase in the number of collisions and radiation interactions between particles, which facilitates the transfer of energy from hot particles to cooler ones, and ultimately results in a decrease in the overall temperature of the plasma.

$N_{C_2}$ values increase with pressure, displaying a diffuse C$_2$ distribution at 25 kPa but a strongly localized distribution at 70 kPa in the top part of the probed zone, down to -120 mm. The $N_{C_2}$ distribution profiles are different from those of $T_{C_2}$ since C$_2$ molecules in the plasma jet are formed by thermochemistry rather than electron impact, resulting in thermodynamically favoured density distributions at specific temperatures. The optimal formation temperature for C$_2$ molecules is around 4,000 K (see section 3.3).

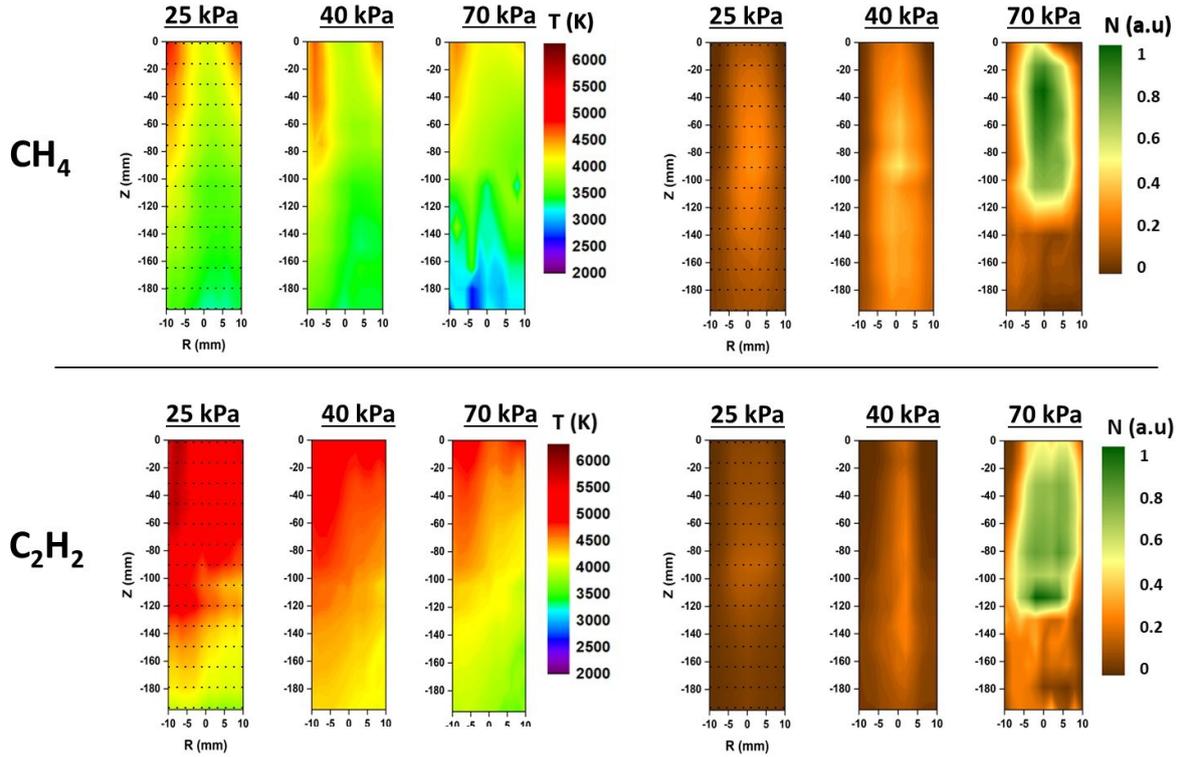

Figure 9. 2D OES maps of $T_{C_2}$ and $N_{C_2}$ using CH$_4$ (1.5 slpm) and C$_2$H$_2$ (1.5 slpm) at different pressures (the points on the maps indicate the location of the optical fibres).

### 3.2.2 Effect of $H/C$ ratio

To investigate the effect of the precursor ratio $H/C$ on plasma characteristics such as $T_{C_2}$ and $N_{C_2}$, we increased $H/C$ to 8 and kept the pressure and input power fixed. Figure 10 compares $T_{C_2}$ distribution of the plasma at a pressure of 45 kPa and a CH$_4$ flow rate of 1.5 slpm and C$_2$H$_2$ of 0.7 slpm. Adding hydrogen, and thus, increasing the $H/C$ ratio, for either C$_2$H$_2$ or CH$_4$ leads to plasma cooling. The rapid quench of the plasma with hydrogen addition is explained by the loss of energy due to the improvement of the thermal conductivity [53,54] and enthalpy of the gas mixture over the temperature range between 3,000 and 6,500 K (Section 3, "Thermodynamic calculations" in the Suppl Info.). However, with both gas CH$_4$ and C$_2$H$_2$ the temperature seems to decrease monotonically down the axis of the reactor. The plasma loses energy through radiation and contact with the atmosphere in the chamber, with no further temperature gain. Increasing $H/C$ ratios from 1 to 8 leads to systematic increase of $N_{C_2}$ with the $N_{C_2}$ maximum at the center of the jet, from -60 mm to -120 mm.

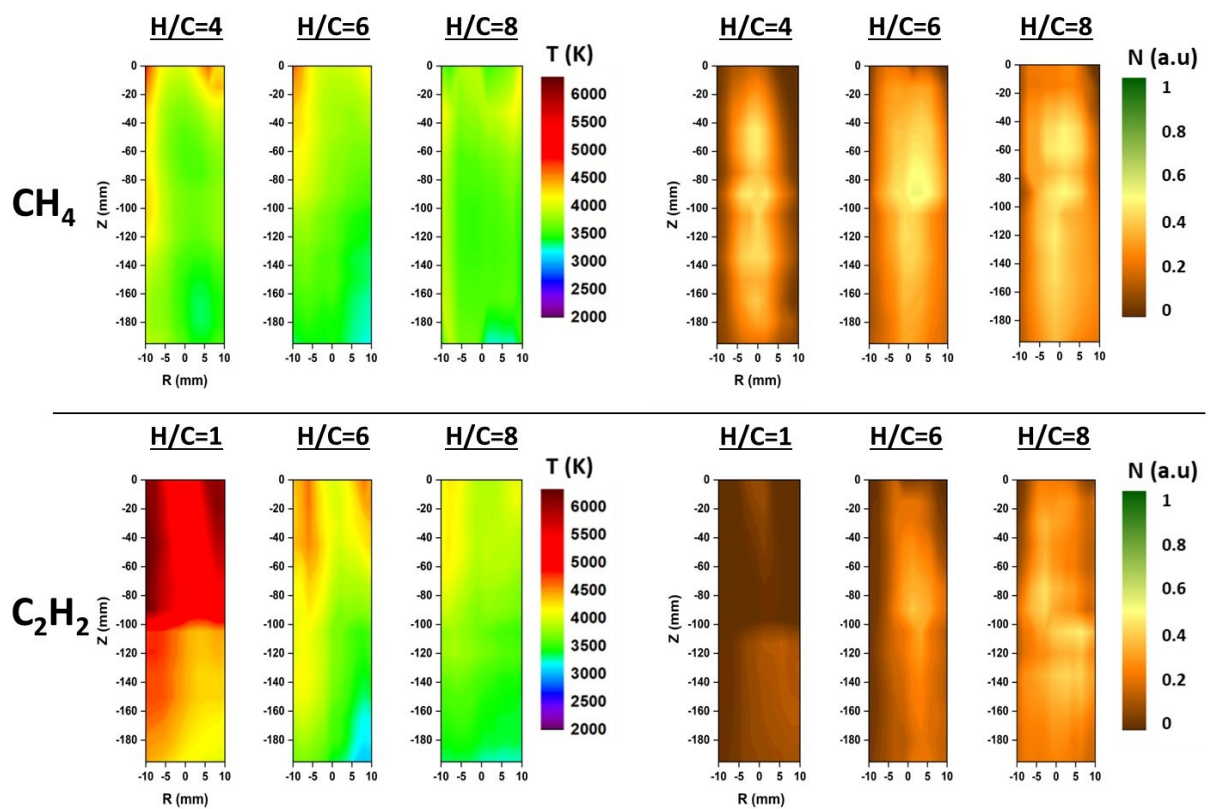

Figure 10. 2D OES maps of $T_{C_2}$ and $N_{C_2}$ at 45 kPa using $CH_4$ (1.5 slpm) and $C_2H_2$ (0.7 slpm) at several $H/C$ ratios.

### 3.2.3 Effect of hydrocarbon flow rate

When the $CH_4$ flow rate is increased from 0.7 to 4 slpm, a significant drop in temperature is observed, as illustrated in Figure 11 (a). This increase in $CH_4$ flow rate leads to a higher concentration of $C_2$ in the plasma jet, which further enhances carbon nanomaterial production, as detailed in Table 2. The higher $CH_4$ flow rate contributes to a decrease in the overall energy density of the plasma, as more energy is consumed by the endothermic $CH_4$ decomposition and therefore cools the plasma, with a negligible impact from forced convection (Section 3, "Thermodynamic calculations" in the Suppl Info.). In addition, Figure 11 (b) shows that the temperature distribution along the injection direction (Z) remains consistent, suggesting that the coefficient for forced convection heat transfer is typically much lower than that associated with the endothermic process of $CH_4$ decomposition.

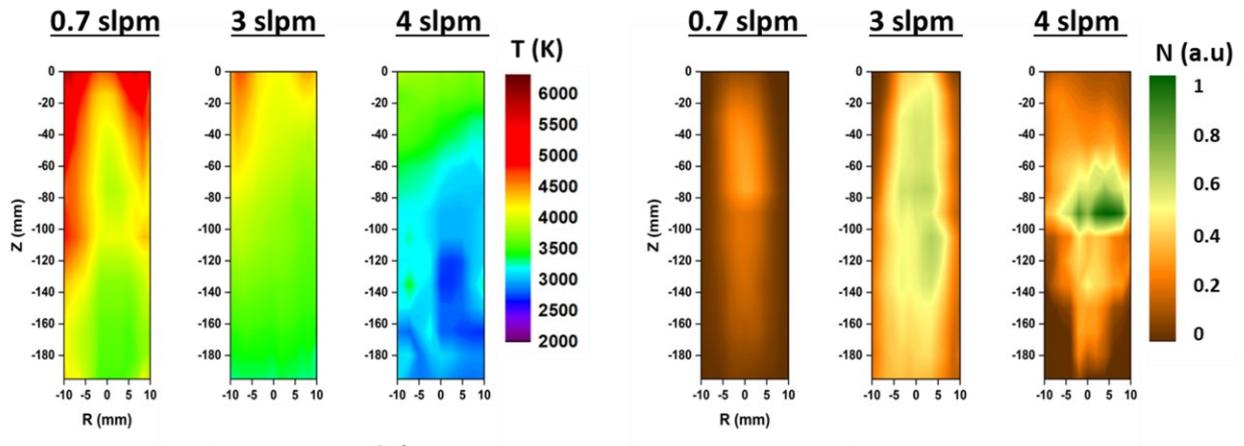

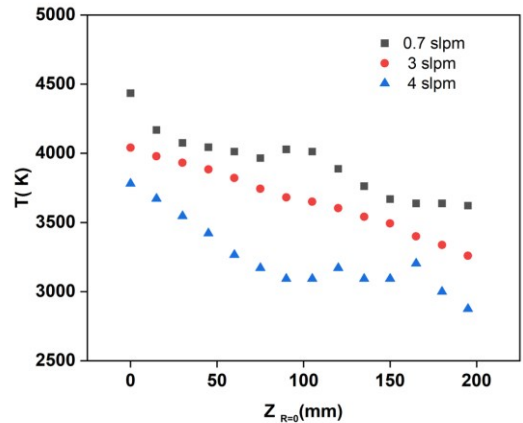

Figure 11. (a) 2D OES maps of $T_{C_2}$ and $N_{C_2}$ at 45 kPa for different CH$_4$ flow rates. (b) Temperature distribution along the injection direction Z at 45 kPa for different CH4 flow rates.

## 3.3 Chemical reactions within the plasma jet

Figure 12 (a, b) shows the mass spectrometer results obtained for various synthesis conditions (1.5 slpm, at 40 kPa and different H$_2$ injections resulting in $H/C = 4, 6, 8$). No signal change was recorded for CH$_4$ after gas injection, which confirms its complete decomposition by the plasma for all the conditions explored. The analysis of the gas composition shows that CH$_4$ is converted not only to H$_2$ along with the carbon nanoparticles but also to C$_2$H$_2$. As no other signal is observed, only H$_2$ and C$_2$H$_2$ are the gaseous products. Adding H$_2$ to CH$_4$ in Ar plasma or increasing the CH$_4$ flow rate results in an increase in H$_2$ and C$_2$H$_2$ produced during the synthesis reaction.

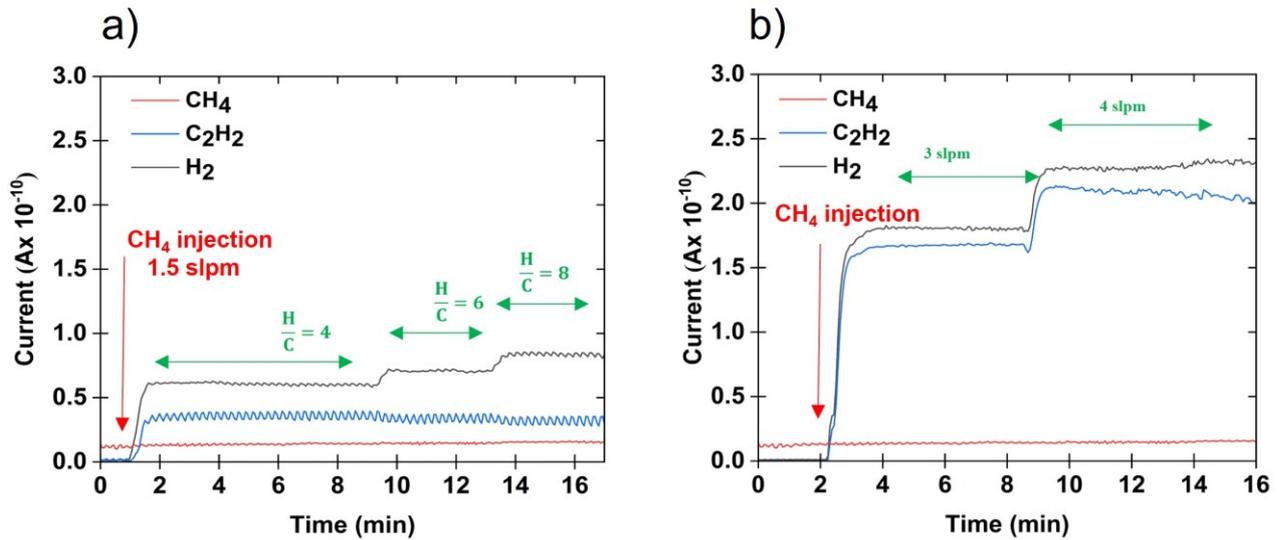

Figure 12. Real-time mass spectrometer data recorded during nanostructure synthesis (a) Using CH$_4$ (1.5) slpm at 40 kPa) and different H$_2$ injections resulting in $H/C$ =4, 6, 8, (b) Using CH$_4$ flow rates of 3 slpm and 4 slpm.

To support the previous observations, the chemical compositions at equilibrium of CH$_4$ and C$_2$H$_2$ in argon between 300 and 8,000 K were simulated with the FactSage 8.2 software. We argue in favour of the assumption of local chemical equilibrium and, therefore, the use of thermodynamic equilibrium calculations with FactSage 8.2. The main arguments are: (i) the complete decomposition of the precursor within the plasma, as confirmed by online mass spectrometry (Figure 12-a, b); (ii) the jet expansion from a narrow nozzle, ensuring a locally homogeneous mixing; (iii) $T_{C_2} = T_{CN}$, implying a local thermal equilibrium between the heavy particles (Section 1.2, "C$_2$ as a plasma thermometric probe" in the Suppl Info.).

When we read diagrams in Figure 13 from right to left (from high to low temperatures), which corresponds in our experiments to the path taken by the species from the top part of the plasma to the bottom, we observe that at high temperatures (> 4,000 K), electrons and ions are present, followed by atomic species H and C. The disappearance of these atomic species is associated with the formation of molecules, particularly C$_2$, which reaches its maximum around 3,800 K. The subsequent disappearance of C$_2$ below 3,500 K is followed by the formation of larger radicals C$_3$, C$_4$, and C$_5$ until 2,800 K, where the formation of solid carbon C(s) is observed, accompanied by the formation of H$_2$ and C$_2$H$_2$. When comparing CH$_4$ and C$_2$H$_2$, both at a gas flow rate of 1.5 slpm, the main distinction is seen in the production of solid carbon C(s) and H$_2$. C$_2$H$_2$ yields more solid carbon than CH$_4$, while CH$_4$ produces more H$_2$ than C$_2$H$_2$. This difference is minimized when hydrogen is introduced and the gas flow rate for C$_2$H$_2$ is reduced to 0.7 slpm. The thermodynamics of the reaction products are comparable, ensuring a uniform environment favourable to the growth of similar nanostructures. TEM images confirm the similarity of GNFs structures regardless of the gas used, as long as the $H/C$ ratio favourable to GNFs is satisfied. The thermodynamic calculations are in full agreement with our mass spectrometry results, where the final products are $H_2$ and $C_2H_2$ alongside solid carbon powder. The findings also align with OES observations where H and C atoms are seen in the upper part of the plasma at temperatures above 4,000 K (Section 1.4, " H and C atomic lines" in the Suppl Info.), and a maximum concentration of $C_2$ is observed between 3,500 and 4,000 K.

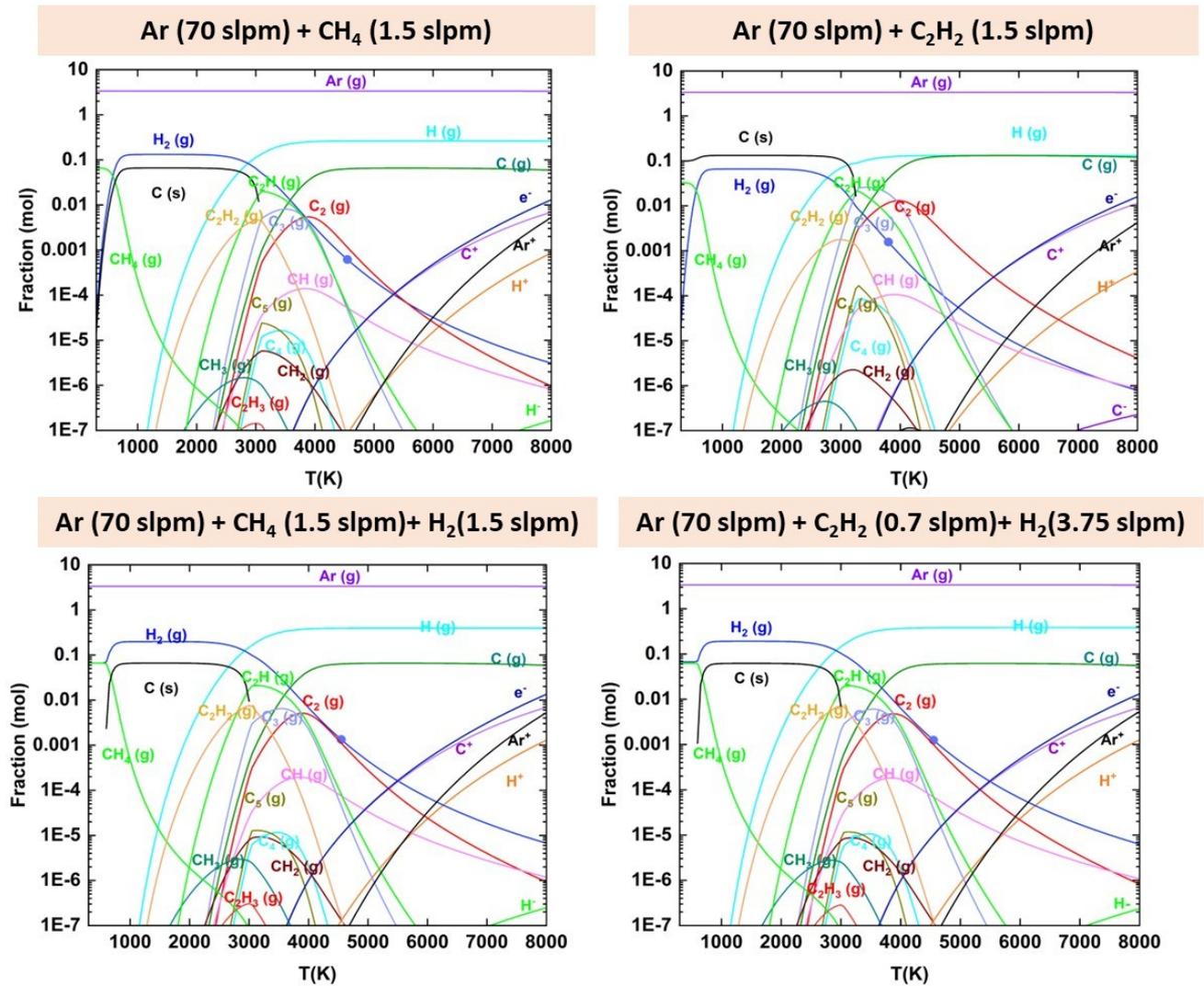

Figure 13. Thermodynamic equilibrium calculations for the systems at 45 kPa: Ar +CH$_4$, Ar +CH$_4$+ H$_2$, Ar +C$_2$H$_2$, and Ar + C$_2$H$_2$ + H$_2$.

# 4 Discussion

**Carbon nanostructure growth in plasma jet**

In this work, we have used *in situ* thermal plasma diagnostics and MS measurements to investigate the formation of carbon nanostructures and complemented the findings by chemical reaction simulations. We have primarily identified graphene nanoflakes, carbon nanohorns, and graphitic nanocapsules within the range of parameters explored. In the following, we discuss the findings to determine the conditions and factors responsible for the formation of these distinctive nanostructures.

At temperatures between 4,000 and 6,000 K, the plasma energy is sufficient to completely decompose the precursors, as evidenced by the complete $CH_4$ conversion (MS results) and the detection of electrons and atomic species like C and H near the torch outlet (Section 1.4, " H and C atomic lines" in the Suppl Info.). The C atoms combine to form $C_2$ molecules in the jet region from -10 to -100 mm below the nozzle with the peak $C_2$ concentrations observed experimentally by OES at temperatures around 3,700 K, similarly to the predicted composition by FactSage 8.2. It is in this region that the nucleation stage occurs, with molecular clusters initiating their formation through bimolecular collisions, ultimately leading to the production of heavier molecules such as $C_3$, $C_4$, and $C_5$. With the continuous addition of adatoms to these clusters, the nucleation gains momentum and nuclei grow. Notably, no powder was produced at pressures lower than 13 kPa nor with carbon source flow rates below 0.5 slpm during the experiments. This observation confirms the need to maintain a $C_2$ concentration threshold to maintain supersaturation.

Based on these observations, the process of carbon nanostructure formation using thermal plasma is reminiscent of the nanoparticle formation observed in aerosol approaches like flame spray pyrolysis and involves two primary stages: nucleation and subsequent growth by condensation [55]. For this process, the transition of gas-phase polycyclic aromatic hydrocarbons (PAHs) such as benzene ($C_6H_6$) and pyrene ($C_{16}H_{10}$) to condensed-phase clusters was perceived as the mechanism driving the nucleation of various carbon nanostructures following the precursor combustion in flames (< 2,000 K) [56–58]. Analogously, this mechanism was then suggested to explain the nucleation of carbon nanostructures in thermal plasma, particularly GNFs [26,59,60]. Nevertheless, this mechanism of nucleation seems unlikely to occur in our case because the plasma temperature in the upper part of the plasma is expected to reach ~6,000 K and, given the flow rates used in this study, should be sufficient to completely decompose the precursors into atomic species. Additionally, our results highlight that the $C_2$ molecules are the fundamental building block for thermal plasma nucleation of carbon nanostructure. Nucleation is also observed at experimental temperatures exceeding 3,500 K, a range not conducive to PAHs formation as confirmed by our thermodynamic calculations.

Particles moving from the nucleation zone enter the growth zone, where the final structure is influenced by factors like residence time in specific temperature ranges. To determine the exact zone of carbon nanostructure formation within the plasma jet, the carbon powder was collected on a stainless steel substrate at different distances from the nozzle exit. Figure 14 (a) shows TEM images of the powder collected on the substrate surface at distances 50 mm and 100 mm below the torch. It reveals that at 50 mm, the nanoparticles were amorphous, while the nanoparticles collected at 100 mm were crystalline GNFs. Given the overall gas flow rate, this suggests that the particles crystallize in less than 10 ms in regions where the temperature is around 3,500 K. This result is consistent with predictions from a recent CFD model calculation from a prior study, which focused on the nucleation and growth of GNFs using thermal plasma [28]. In their findings, they estimated the nucleation and growth processes to occur within 14-19 ms at temperatures ranging from 3,000 K to 5,000 K.

In addition to local carbon density and residence time, the final morphology of nanoparticles also depends largely on the temperature gradient and quenching rate [61]. We note that carbon nanomaterials produced by thermal plasma are typically smaller than nanocarbons produced by other plasma-related methods, such as PECVD. We can attempt here an explanation, based on the difference in the transport phenomena exclusive to the thermal plasma processes. We start by estimating the vertical temperature gradient between $Z = 0$ mm and $Z = -195$ mm along the center line at a radial position of $R = 0$ ($\nabla T_Z^{R=0}$). Figure 14 (b) shows the correlation between $\nabla T_Z^{R=0}$ and the average size of the nanostructures. Once formed, the nanoparticles are easily transported away from the growth region by two phenomena: convection and diffusion. We also expect thermophoresis diffusion [62] to drive transport in the radial direction toward the cooled reactor walls. As a result of the larger temperature gradient, the residence time of particles in this zone is shorter, leading to smaller particles. Finally, a variety of stable carbon nanostructures were identified in this study including GNFs, CNHs, and onions like GNCs. These different nanostructures were found to be selectively produced depending on the reaction conditions.

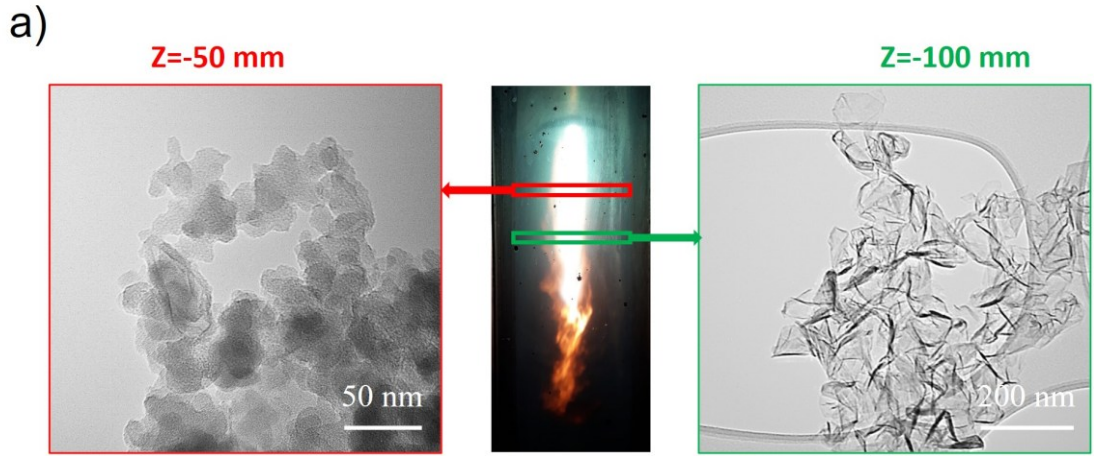

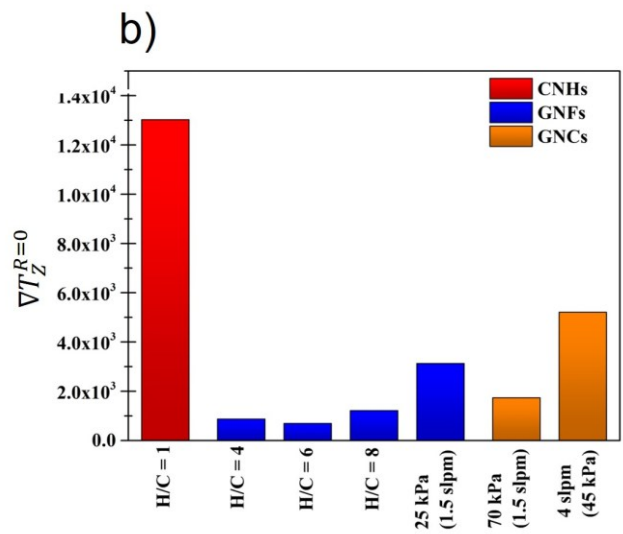

Figure 14. (a) Digital picture of the $CH_4$/Ar plasma jet and TEM images of the nanostructures synthesized within it at -50 mm and -100 mm from the nozzle exit (using $CH_4$ of 1.5 slpm at 45 kPa). (b) Temperature gradient at $R = 0$ mm for various synthesis conditions with the colour corresponding to the nanostructures obtained.

**Graphene nanoflakes**

In this study, we have demonstrated that the formation of GNFs is intricately linked to temperature, hydrogen addition and carbon content in the plasma. GNFs, characterized by a planar structure, predominantly nucleate at temperatures exceeding 3,500 K. The role of hydrogen is crucial in the selectivity of GNFs within a thermal plasma jet. The GNFs are synthesized predominantly with $H/C > 1$, either with $CH_4$, $CH_4+H_2$, or $C_2H_2+H_2$. During the nucleation stage, hydrogen aids in removing dangling carbon bonds from the edge of GNFs, forming carbon-hydrogen (C-H) bonds [63,64]. This promotes the lateral growth of the graphene flakes and prevents the edges from curling or closing, thereby avoiding the formation of closed carbon structures. However, at an $H/C$ ratio of 1, there is insufficient hydrogen to stabilize GNFs growth, leading to closed forms of carbon such as bud-like CNHs. In the later stages of growth, despite the initial anti-curling influence of hydrogen, GNFs exhibit bent or scrolled edges. This feature is attributed to microscopic wrinkling, which provides thermodynamically stable three-dimensional structures at localized regions, contrasting with the earlier unstable planar configurations [65].

High $CH_4$ flow rates are not optimal for GNFs formation, even with elevated $H/C$. Based on FactSage 8.2 simulations and the OES data, the nucleation of GNFs is thermodynamically allowed within a temperature range of 3,500-4,500 K and pressure range of 25-45 kPa, and chemically favourable at a ratio of $H/C > 1$, but kinetically limited by a relatively low $C_2$ density. An increase in the $CH_4$ flow rate directly elevates the density of $C_2$ in the plasma. This heightened density subsequently affects the collision frequencies, playing a key role in determining the resulting carbon morphologies. Notably, reports indicate that high collision frequency results in three-dimensional nuclei with a nearly isotropic growth site, favouring GNCs formation and other graphitic-like structures [28].

**Graphitic nanocapsules**

We confirm that a high concentration of $C_2$ is essential to the formation of GNCs. Using OES and TEM, we determined that increased pressure and carbon precursor flow rates in the plasma synthesis process directly correlate with a higher local concentration of $C_2$ molecules, facilitating the formation of GNCs. Hydrogen plays a critical role in the selectivity of GNCs at high pressure. For instance, the presence of $C_2H_2$ ($H/C = 1$) at 70 kPa consistently led to the synthesis of carbon black structures instead of GNCs. We put forward that GNCs with a polyhedral form and concentric multilayer graphitic shells are an evolved form of GNFs. As detailed in our previous work on the synthesis and growth of GNCs [25], we proposed two possible mechanisms of nucleation and growth responsible for the formation of GNCs: one involving the curling of thin GNFs and another involving the interfacial delamination of GNFs. HRTEM showed that the external layers of the GNCs structure grew through the epitaxial addition of carbon adatoms to specific crystal (002) facets of the GNCs shell [25].

**Carbon nanohorns**

CNHs in the form of bud-like structures were observed using $C_2H_2$ ($H/C = 1$) as the precursor. Hydrogen seemed to favour the synthesis of GNFs, while recent studies by Casteignau et al. [3,66] (using the same ICP process, but with a lower argon flow rate) found that hydrogen had the opposite effect on CNHs and improved their yield. In Casteignau's study, it was found that the presence of a promotor gas, particularly hydrogen or nitrogen, played a crucial role in CNHs growth, although the underlying mechanisms remained unclear. This suggests that the $H/C = 1$ ratio is not necessarily the key factor in the selectivity of CNHs synthesis. Previous literature on CNHs synthesis methods, such as DC arc [67], laser ablation [68], and joule heating [69], that do not require a catalyst, involve a carbon-rich source (such as graphite), temperatures above 3,200 K, an inert environment (such as argon or helium), power between 5-30 kW, and pressures less than 100 kPa. These conditions are close to the ICP thermal plasma environment used in this study. Molecular-dynamics simulations have suggested that a rapid quench at a temperature above 2,000 K is key to favouring CNHs synthesis using hydrogen-free graphitic layers [70]. In this study, the plasma jet environment achieved higher plasma core temperatures with $C_2H_2$ at 45 kPa compared to $CH_4$, resulting in a more severe quench. The primary factor influencing CNHs structures appears to be the temperature gradient associated with this rapid quench. Given their size of 50 nm, CNHs likely exit the jet rapidly due to a strong gradient of temperature, limiting their growth or transformation into other structures [61]. However, the exact mechanism by which the horns are formed is still unknown. We are currently investigating this mechanism, and it will be the focus of our ongoing research.

## 5 Conclusion

In this study, utilizing an ICP torch, we produced various carbon nanostructures, notably graphene nanoflakes (GNFs), carbon nanohorns (CNHs), and graphitic nanocapsules (GNCs). Using a combination of extensive materials characterization techniques and *in situ* optical emission spectroscopy, we identified the relationships between process parameters, plasma characteristics, and nanostructure morphology, essential for precise control over carbon nanostructure growth in a thermal plasma environment. Our principal findings can be summarized as follows:

- While the type of precursor, whether it is $CH_4$ or $C_2H_2$, was a determinant factor for nanostructure type, maintaining a uniform $H/C$ atomic ratio consistently yielded a predictable nanostructure.
- Gas flow rate, pressure, and H/C ratio played a crucial role in the nanocarbon morphology through the control of the local $C_2$ radicals concentration. High $C_2$ densities favour the formation of compact structures such as GNCs, while lower densities favour the emergence of less dense structures, such as GNFs and CNHs.

- Temperatures in the jet nucleation zone exceeded 3,500 K and the crystal growth window was located below ~50 and 100 mm from the torch nozzle. This highlights the importance of the temperature variations and residence time in this zone for the nanocarbon morphology and crystallinity.

This in-depth study provides a better understanding of the complex interaction of parameters in the synthesis of carbon nanostructures by thermal plasma. By adjusting parameters such as precursor type, gas flow rate and pressure, it is possible to exert control over the desired nanostructures. The results offer a refined perspective, particularly on the role of $C_2$ radicals, establishing a basis for more predictable synthesis.

## Acknowledgment

The authors acknowledge funding from the Green Surface Engineering for Advanced Manufacturing (Green-SEAM), a Strategic Network funded by the Natural Sciences and Engineering Research Council of Canada (NSERC), the Canada Research Chair Program, and Université de Sherbrooke. The authors also appreciate the technical support of Dr. Kossi Béré from the Plasma Process and Integration of Nanomaterials lab. We would like to thank Charles Bertrand from the Plateforme de Recherche et d'Analyse des Matériaux (PRAM) of Université de Sherbrooke for help in the acquisition of data related to material characterization.